\documentclass[%
reprint,
superscriptaddress,
amsmath,amssymb,
aps,
prb,
floatfix,
]{revtex4-2}

\usepackage{graphicx}
\usepackage{dcolumn}
\usepackage{booktabs}
\usepackage{bm}
\usepackage{braket}
\usepackage[colorlinks]{hyperref}
\usepackage{color}
\usepackage[usenames,dvipsnames]{xcolor}
\usepackage{comment}
\usepackage[export]{adjustbox}
\usepackage{siunitx}
\usepackage{fouriernc}

\emergencystretch=\maxdimen
\newcommand{\be}{\begin{equation}}
\newcommand{\ee}{\end{equation}}

\begin{document}
	
	\small
	
	\title{1/{\itshape f} frequency fluctuations due to kinetic inductance in CoSi\textsubscript{2} microwave cavities}

     \author{Weijun Zeng}
	\affiliation{Low Temperature Laboratory, Department of Applied Physics, Aalto University, P.O. Box 15100, FI-00076 Espoo, Finland}

	\affiliation{QTF Centre of Excellence, Department of Applied Physics, Aalto University, P.O. Box 15100, FI-00076 Aalto, Finland}
 
	\author{Ilari Lilja}
	\affiliation{Low Temperature Laboratory, Department of Applied Physics, Aalto University, P.O. Box 15100, FI-00076 Espoo, Finland}
	
	\affiliation{QTF Centre of Excellence, Department of Applied Physics, Aalto University, P.O. Box 15100, FI-00076 Aalto, Finland}
 
	\author{Ekaterina Mukhanova}
	\affiliation{Low Temperature Laboratory, Department of Applied Physics, Aalto University, P.O. Box 15100, FI-00076 Espoo, Finland}
	
	\affiliation{QTF Centre of Excellence, Department of Applied Physics, Aalto University, P.O. Box 15100, FI-00076 Aalto, Finland}%

    \author{Elica Anne Heredia}
    \affiliation{International College of Semiconductor Technology, National Yang Ming Chiao Tung University, Hsinchu 30010, Taiwan}

	\author{Chun-Wei Wu} 
    \affiliation{Department of Electrophysics, National Yang Ming Chiao Tung University, Hsinchu 30010, Taiwan}

    \author{Juhn-Jong Lin}
	\affiliation{Department of Electrophysics, National Yang Ming Chiao Tung University,
Hsinchu 30010, Taiwan}

 \author{Sheng-Shiuan Yeh}  \email{ssyeh@nycu.edu.tw}
	
	\affiliation{International College of Semiconductor
Technology, National Yang Ming Chiao Tung University, Hsinchu 30010, Taiwan}

	\affiliation{Center for Emergent Functional Matter Science, National Yang Ming Chiao Tung University, Hsinchu 30010, Taiwan}

	\author{Pertti Hakonen}  \email{pertti.hakonen@aalto.fi}
	\affiliation{Low Temperature Laboratory, Department of Applied Physics, Aalto University, P.O. Box 15100, FI-00076 Espoo, Finland}
	
	\affiliation{QTF Centre of Excellence, Department of Applied Physics, Aalto University, P.O. Box 15100, FI-00076 Aalto, Finland}

	\date{\today}
	
	\begin{abstract}		
 Cobalt disilicide provides a promising nearly-epitaxial superconducting material on silicon, which is compatible with high-density integrated circuit technology. We have characterized CoSi\textsubscript{2} superconducting microwave cavities around \qty{5.5}{GHz} for resonance frequency fluctuations at temperatures \qtyrange[range-phrase = --,range-units = single]{10}{200}{mK}. We found relatively weak fluctuations $(\delta f/f)^2$ following the spectral density $A/f^{\gamma} $, with $A \simeq 6 \times 10^{-16}$ and $\gamma$ slightly below 1 at an average number of photons of $10^4$; the noise decreased with measurement power as $1/P^{1/2}$. We identify the noise as arising from kinetic inductance fluctuations and discuss possible origins of such fluctuations.
	\end{abstract}
	
	\maketitle
	

	\section{Introduction}

 Low noise materials working at microwave frequencies are essential for nano- and micro-fabrication of quantum circuits \cite{deLeon2021}. Besides the noise properties at low frequencies and microwave frequencies \cite{Lisenfeld2019,Wang2020,Falci2023}, compatibility with existing large scale integrated circuit technologies is an important issue for selection of future materials \cite{Tolpygo2016}. Unfortunately, proper understanding of microscopic noise mechanisms in the employed superconducting materials, such as aluminum, is not fully understood. There is, however, ample evidence that surfaces and interfaces act as important sources of noise. Consequently, epitaxial systems form an important class of materials whose potential has not yet been completely exhausted. 
 
 In this work, we characterize low frequency noise properties of one promising nearly-epitaxial material, cobalt dicilicide (CoSi\textsubscript{2}), as a platform for GHz frequency low-loss microwave cavities. Superconducting CoSi\textsubscript{2} is a high-quality material that can be manufactured using a simple self-limiting process: a layer of cobalt of thickness $t_{Co}$ is annealed on top of silicon, which yields a $3.5t_{Co}$-thick CoSi\textsubscript{2} conductor embedded into Si \cite{Chen2004}. This embedded CoSi\textsubscript{2} structure has an excellent lattice match to silicon. Owing to the nearly-epitaxial match, the CoSi\textsubscript{2}/Si interface is nearly defect-free, which results in an exceptionally low level of $1/f$ resistance noise \cite{Chiu2017}.

 Thin CoSi\textsubscript{2} films possess a substantial kinetic inductance \cite{Mukhanova2023}. This inductance scales with normal state resistance, i.e. inversely with the cross-sectional area of the conductor. One of the questions that we aim to answer in our present work is whether the small $1/f$ resistance noise $\delta R^2/R^2$ in the normal state turns into small $1/f$ noise in the kinetic inductance in the superconducting state, i.e. into ultra small fluctuations in the resonant frequency of a microwave cavity.

 The $1/f$ resistance noise $\delta R^2/R^2$ is due to parameter fluctuations \cite{Kamada2023} and it can be described in normal conductors as $
 \frac{\delta R^2}{R^2}=\frac{S_I}{I^2}=A\frac{I^{\beta}}{f^{\gamma}}$, where the power spectrum of current $I$ is denoted by $S_I$ at constant voltage bias, $\beta \simeq 0$ 
 and $ \gamma \simeq 1$. Basic models for $1/f$ noise can be found in Ref.~\cite{Kogan2008,Fleetwood2015,Grasser2020} and summarized in Ref.~\cite{Falci2023}. Most often, $1/f$ noise models are based on the summation of an ensemble TLSs \'a la McWorther \cite{Burstein1957}, but more involved theories based on impurity clustering have also been proposed \cite{Pelz1987,Giordano1989,Kamada2021}.

 In superconductors, the basic parametric variation is noise in kinetic inductance $\delta L_\mathrm{K}^2/L_\mathrm{K}^2$, which becomes prominent in metals and alloys that have a large normal state resistance, such as in thin CoSi\textsubscript{2} films. In superconducting microwave cavities, besides fluctuations in $L_\mathrm{K}$, noise in the dielectric permittivity of the substrate leads to fluctuations in the resonant frequency. Customarily, these fluctuations in the cavity frequency are expected to arise from two level states (TLS) active around the resonant frequency \cite{Zmuidzinas2012,Lisenfeld2019,Wang2020}.  Such low-frequency dielectric (or inductive) noise turns into time dependence of the circuit parameters leading to unwanted detuning of narrow band circuits, such as qubits connected to their read-out cavities. A generic feature of the TLS noise is the reduction of observed noise with increasing measurement power, typically as $P^{\alpha}$ with $\alpha\sim -0.5$.
 

Because fluctuations of the resonant frequency in our CoSi\textsubscript{2} cavity are expected to be small, extra care was taken to avoid low frequency noise from the measurement system itself. We employed a Pound locking scheme as our starting point \cite{Pound1946}, which is known to avoid $1/f$ noise issues present in homodyne measurement setups. However, we did not use the setup in the closed loop (locked) regime, choosing instead to operate in the open loop regime, where it is easier to achieve a larger bandwidth. Moreover, as the overall drift is small in our samples, it is quite easy to remain in the linear error regime for extended periods without the need to adjust the operation frequency. In our Pound locking setup, the frequency fluctuations are obtained from the error voltage $\varepsilon$ according to \cite{Lindstrom2011}  
 \be
\frac{S_\varepsilon}{f_0^2}\left(\frac{\mathrm{d}f_\mathrm{r}}{\mathrm{d}\varepsilon}\right)^2 = \frac{S_{f_\mathrm{r}}}{f_0^2} = A\frac{P_\mathrm{m}^\alpha}{f^\gamma},
\label{eq:1_f}
 \ee 
where $S_{\varepsilon}$ and $S_{f_{\mathrm{r}}}$ denote the power spectral density of $\varepsilon$ and the fluctuating cavity frequency $f_{\mathrm{r}}$, respectively, and $f_0$ is the mean of the cavity resonant frequency. In the last form, $P_\mathrm{m}$ refers to measurement power, and its influence is described by the exponent $\alpha$. 


\section{Experimental methods}

\begin{figure}
    \centering
    \includegraphics[width=\linewidth]{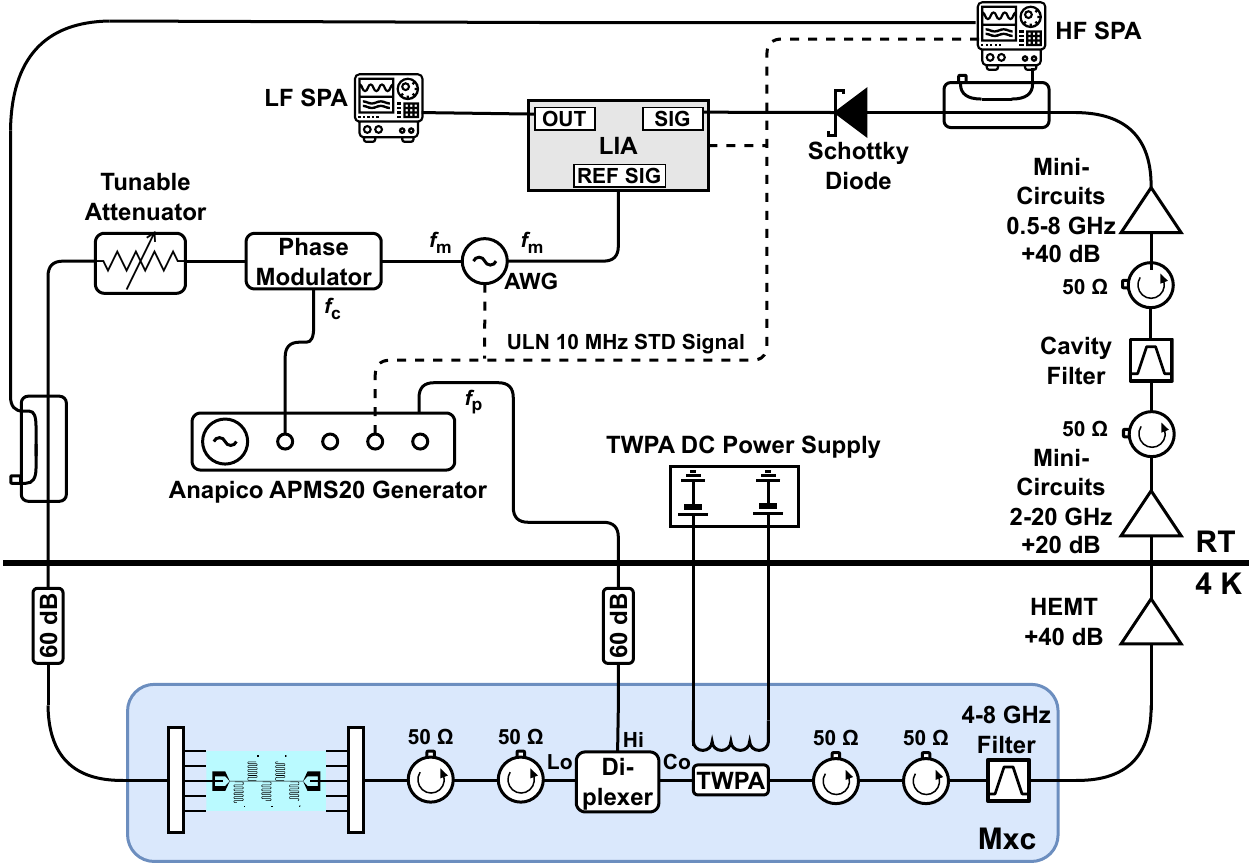}
    \caption{\textbf{Schematics of the Pound-locking system}, employed in the open-loop detection circuit. The CoSi\textsubscript{2} cavity is connected via a microwave switch, two circulators, and a diplexer to the TWPA. The amplified carrier wave around 6~GHz (on and off resonance) is guided to a 4~K HEMT amplifier using two \qtyrange[range-phrase = --,range-units = single]{4}{12}{GHz} circulators and a \qtyrange[range-phrase = --,range-units = single]{4}{8}{GHz} band pass filter.
    }
    \label{fig:setup}
\end{figure}

 Our experimental setup is illustrated in Fig.~\ref{fig:setup}. It is based on a standard Pound locking scheme operating at microwave frequencies. We have employed both closed loop and open loop operations. The experiments with feedback connected yielded equivalent results as the open loop configuration \footnote{A drawback of our PID controller was that it gives non-constant intervals between the points. Consequently, Lomb-Scargle Periodogram analysis had to be performed instead of the regular Fast Fourier Transform. With such analysis, agreement between open and closed loop measurements was achieved}.

The cavity is connected through a six-port microwave switch and two circulators to a traveling wave amplifier based on SNAIL unit cells \cite{Ranadive2022} and operating between \qtyrange[range-phrase = --,range-units = single]{4}{8}{GHz} frequency band. The TWPA is rf pumped at a frequency around 11~GHz via a diplexer in which the other input port passes the signal frequency band from the cavity. The operating point was tuned to minimize the intermodulation products in the traveling wave parametric amplifier (TWPA) output signal by tuning the battery-driven flux bias \cite{Lilja2024}. The amplified signal from the cavity, after further RT amplification and filtering, is connected to a detection diode and to a microwave spectrum analyzer via a directional coupler, which we employ for signal monitoring purposes. The zero-bias Schottky diode detector (Krytar 203BK) is connected to an rf frequency lock-in amplifier (LIA, Stanford Research SR844) locked to the modulation frequency $f_\mathrm{m}$ around 2~MHz from the arbitrary waveform generator (AWG, Agilent 33500B); this high modulation frequency, which eliminates problems with low frequency noise of typical homodyne detection schemes, is one of the attractive features of the Pound locking scheme. The LIA provides the error signal (see Eq.~(\ref{eq:error}) below) that is analyzed using a low frequency spectrum analyser (Stanford Research SRS785).

In the open loop mode, the modulated drive signal, containing the carrier frequency $f_\mathrm{c}$ ($f_\mathrm{c}\approx f_0$) and side bands $f_\mathrm{c} \pm f_\mathrm{m}$, with nearly equal voltage amplitudes, is produced using a high-quality microwave generator (XCO, Anapico APMS20) and the AWG. The modulation frequency is generated by the AWG, which drives the phase modulator unit (Minicircuits ZX05-83+) for generating the sidebands $f_\mathrm{c} \pm f_\mathrm{m}$ from the carrier signal of the XCO. The modulated signal is fed to the cavity through a programmable attenuator (Vaunix LDA-602) and a microwave switch. The time base of the AWG is locked to the 10~MHz reference signal derived from the XCO, which contains an Ultra Low Noise (ULN) reference option.

Our CoSi\textsubscript{2} samples are notch-type cavities with five resonators coupled to a central read-out line \cite{Mukhanova2023}. Transmission $S_{21}(f_\mathrm{r})$ for a notch type resonator with good phase and amplitude calibration can be written as \cite{Gao2008,Khalil2012,Weides2015}
\begin{equation}
S_{21}(f_\mathrm{p}) = \left( 1 - \frac{\left(Q_\mathrm{T}/|Q_\mathrm{cpl}|\right)\mathrm{e}^{\mathrm{i}\varphi} }{ 1 + 2\mathrm{i}\delta Q_\mathrm{T} } \right)
\label{eq:notch}
\end{equation}
where $Q_\mathrm{T}$ is the loaded quality factor, $|Q_\mathrm{cpl}|$ the quality factor related to the coupling to external circuit, $\varphi$ quantifies impedance mismatch, $f_\mathrm{p}$ the probing frequency, and $\delta=(f_\mathrm{p}-f_\mathrm{0})/f_\mathrm{0}$ denotes the normalized frequency shift from the resonance frequency $f_0$. Employing open source fitting tools \cite{resonatortools},
it is found, that Eq.~(\ref{eq:notch}) agrees perfectly to our measured data as demonstrated in Ref.~\onlinecite{Mukhanova2023}. The internal quality factor $Q_{\mathrm{int}}$, in turn, can be obtained from the fitted values of $Q_{\mathrm{cpl}}$ and $Q_{\mathrm{T}}$ using $Q_{\mathrm{int}}^{-1}=Q_{\mathrm{T}}^{-1}-Q_{\mathrm{cpl}}^{-1}$. In our cavities with $d=25$ and 100~nm, fitting of Eq.~(\ref{eq:notch}) to the data indicates that coupling to external circuit is quite strong, and $Q_{\mathrm{int}} \sim Q_{\mathrm{cpl}}$, while $Q_{\mathrm{int}}$ dominates in 10~nm CoSi\textsubscript{2} films. 
 
The voltage that enters the diode detector is a sum of three frequency components that have traveled through the notch-type resonator. Under symmetric side-band response, the phase response is given by an error signal of form
\footnote{ 
In the absence of phase shift between side band frequencies $f_0 \pm f_\mathrm{m}$, the imaginary part is given by $-\Im{S_{21}}(f_0-f_\mathrm{m}) \Re{S_{21}}(f_0)-\Im{S_{21}}(f_0+f_\mathrm{m}) \Re{S_{21}}(f_0)+\Im{S_{21}}(f_0) \Re{S_{21}}(f_0-f_\mathrm{m}) + \Im{S_{21}}(f_0) \Re{S_{21}}(f_0+f_\mathrm{m})$, while the real part is
$-\Im{S_{21}}(f_0) \Im{S_{21}}(f_0-f_\mathrm{m})+\Im{S_{21}}(f_0) \Im{S_{21}}(f_0+f_\mathrm{m})-\Re{S_{21}}(f_0) \Re{S_{21}}(f_0-f_\mathrm{m}) + \Re{S_{21}}(f_0) \Re{S_{21}}(f_0+f_\mathrm{m})$. Under symmetric sideband response, the real part vanishes and we are left with the error signal in the imaginary part.}.
Since $\Re{S_{21}(f_0 \pm f_\mathrm{m})} \simeq 1$, 
the error signal for side bands without phase shift can be approximated by
\begin{equation} \label{eq:error}
\varepsilon=2\Im{ \left( 1 - \frac{\left(Q_\mathrm{T}/|Q_\mathrm{cpl}|\right)\mathrm{e}^{\mathrm{i}\varphi} }{ 1 + 2\mathrm{i}\delta Q_\mathrm{T} } \right)\simeq \frac{4Q_\mathrm{T}^2\cos\varphi}{Q_\mathrm{cpl}}\delta - \frac{2Q_\mathrm{T}\sin\varphi}{Q_\mathrm{cpl}}}
\end{equation}
where terms of the order of $\propto \delta^2$ and higher have been omitted. As seen from the second term on the right side of Eq.~(\ref{eq:error}), impedance mismatch produces an offset in the error function. Therefore, the slope of the error function $k=\mathrm{d}\varepsilon/\mathrm{d}\delta$ has to be recorded separately for all measurement conditions to verify the linear operation regime; $k$ is the conversion factor that converts measured noise to actual frequency fluctuations. In the error function measurements, the phase shift between frequencies $f_0 - f_\mathrm{m}$ and $f_0 + f_\mathrm{m}$ was adjusted approximately to \qty{180}{\degree} by tuning $f_\mathrm{m}$ to a proper value. However, due to the impedance mismatch and as per the offset, the exact value isn't simple to distinguish. 
	
\section{Experimental results}
The designed resonance frequency for the studied $\lambda/4$ cavities with length $L_\mathrm{line}=$~\qtyrange[range-phrase = --,range-units = single]{3.86}{4.36}{mm} was $f_0^\mathrm{dgn}=$~\qtyrange[range-phrase = --,range-units = single]{6.83}{7.73}{GHz}. The actual measured resonance frequency for 25-nm-thick cavities was $f_0=$~\qtyrange[range-phrase = --,range-units = single]{5.21}{5.88}{GHz}, while the internal quality factors amounted to $Q_\mathrm{int} =$~\numrange[range-exponents = combine,range-phrase = --,range-units = single]{1.0e4}{2.1e4} at base temperature of the cryostat $T\simeq$ \qty{10}{mK}. The nominal characteristic impedance of 25-nm samples was $Z_0=\sqrt{L/C}=$~\qty{50}{\ohm}, but due to kinetic inductance, the actual value was closer to $Z_0=$~\qty{65}{\ohm}. The superconducting transition temperature was found to be at $T_\mathrm{c} =$~\qty[separate-uncertainty,separate-uncertainty-units = single]{1.34(0.03)}{K}, whereas the sheet resistances $R_\mathrm{s}$ at 1.5~K were 7.1, 2.4, and \qty{0.24}{\ohm} for the CoSi\textsubscript{2} films with thicknesses of $d=10$, 25, and 105~nm respectively. For further details, we refer to Ref.~\onlinecite{Mukhanova2023}.

\begin{figure*}[hbt]
    \centering
    \includegraphics[width=\linewidth]{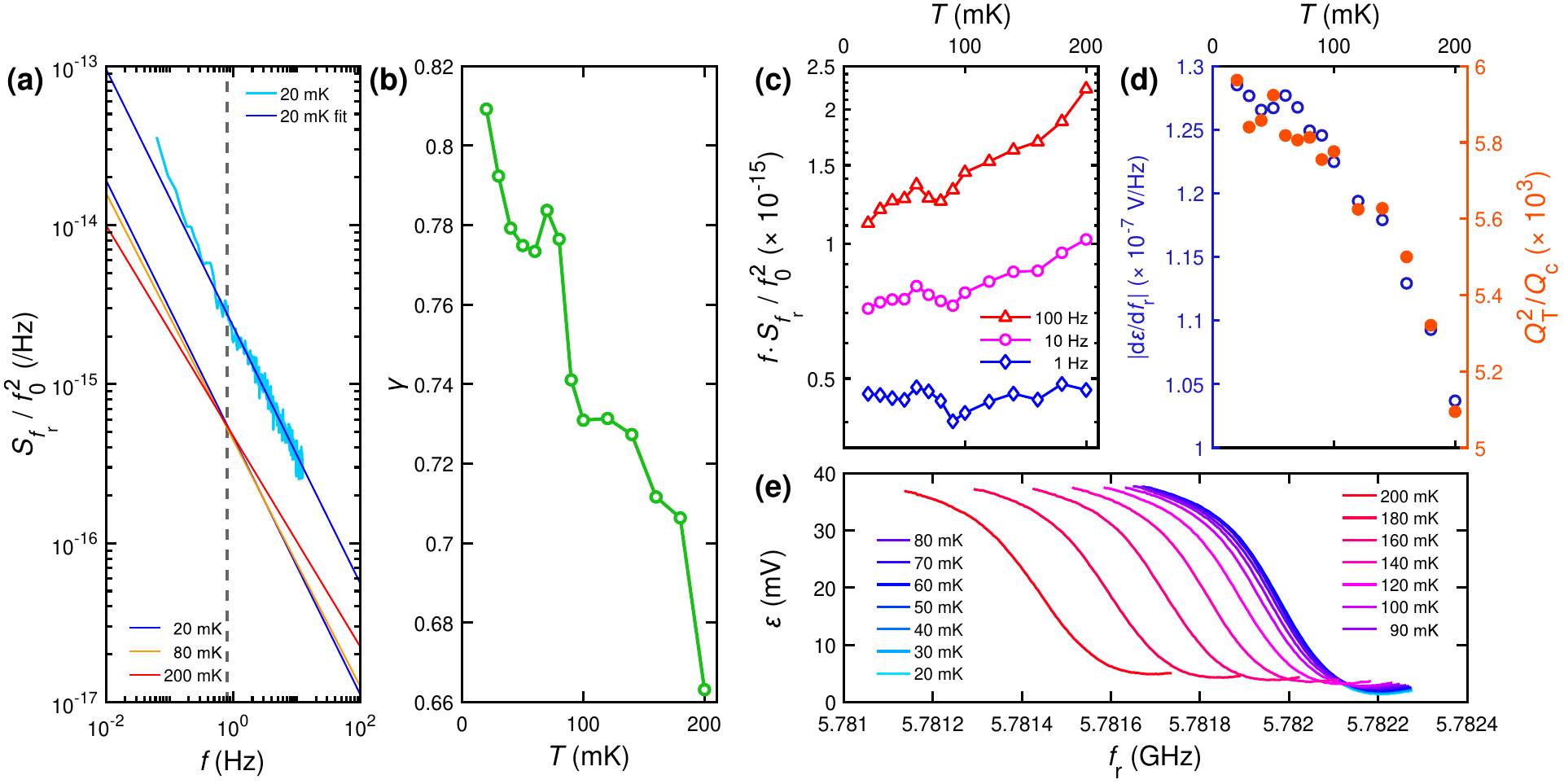}
    \caption{
    {\bf Frequency noise measured at different temperatures} with fixed carrier power {\itshape P}\textsubscript{m} = --100~dBm (average photon number $\bar{n}=10^4$).
    {\bf(a)} Noise power spectra fitted using Eq.~(\ref{eq:1_f}) for the 25-nm-thick CoSi\textsubscript{2} film at temperatures indicated in the legend. A raw noise spectrum with its fit are also included, which are shifted upwards by a multiplier of five for enhanced visibility. 
    {\bf(b)} Exponent $\gamma$ as a function of temperature, determined by fitting Eq.~(\ref{eq:1_f}) to the data. The measured data has been limited to $T<0.2$~K due to temperature fluctuations starting to play an increasing role in noise fluctuations at $T > 0.2$~K.
    {\bf(c)} Noise power as a function of temperature at 1, 10 and \qty{100}{Hz} taken from fits to measured noise spectra.
    {\bf(d)} Slope of the error function (Eq.~(\ref{eq:error})) and $Q$-factor as function of temperature.
    {\bf(e)} Experimentally determined error function of a 25-nm-thick notch-type CoSi\textsubscript{2} microwave cavity measured at temperatures from 20 to 200~mK (from right to left).
    }
    \label{fig:NoisePower_T}
\end{figure*}

A few fits of the measured power spectra for frequency fluctuations at different temperatures between \qtyrange[range-phrase = --,range-units = single]{20}{200}{mK} are displayed in Fig.~\ref{fig:NoisePower_T}(a). The upper temperature limit of 200~mK was chosen as the cavity frequency became quite sensitive to temperatures owing to the strong temperature dependence of cavity frequency caused by the large kinetic inductance of the employed superconducting material. Consequently, the measured low frequency noise started having extra variation owing to variations in the temperature stability during the data collection period.

Several measured error curves of our notch-type resonator with $d=25$~nm are displayed in Fig.~\ref{fig:NoisePower_T}(e); the corresponding measured $S_{21}(\omega)$ can be found in Ref.~\cite{Mukhanova2023}. A large suppression of transmission at resonance indicates that this device is nearly critically coupled with $\frac{Q_\mathrm{T}}{Q_\mathrm{cpl}} \sim \frac{2}{3}$. Because we are interested in noise at small power levels at the carrier frequency, our experiments were performed using a low-noise TWPA as a preamplifier \cite{Perelshtein2022}. Pound-locking-type modulation scheme in conjunction with the TWPA allowed us to measure $1/f$ noise efficiently at powers corresponding to \num{\sim 100} quanta in the cavity. In the course of our experiments, we also investigated the noise induced by the TWPA \cite{Lilja2024}. We found that the operating point of the TWPA had to be carefully tuned to find a sweet operating spot for low $1/f$ noise performance \cite{Lilja2024}. 

 
\begin{figure}
    \centering
    \includegraphics[width=\linewidth]{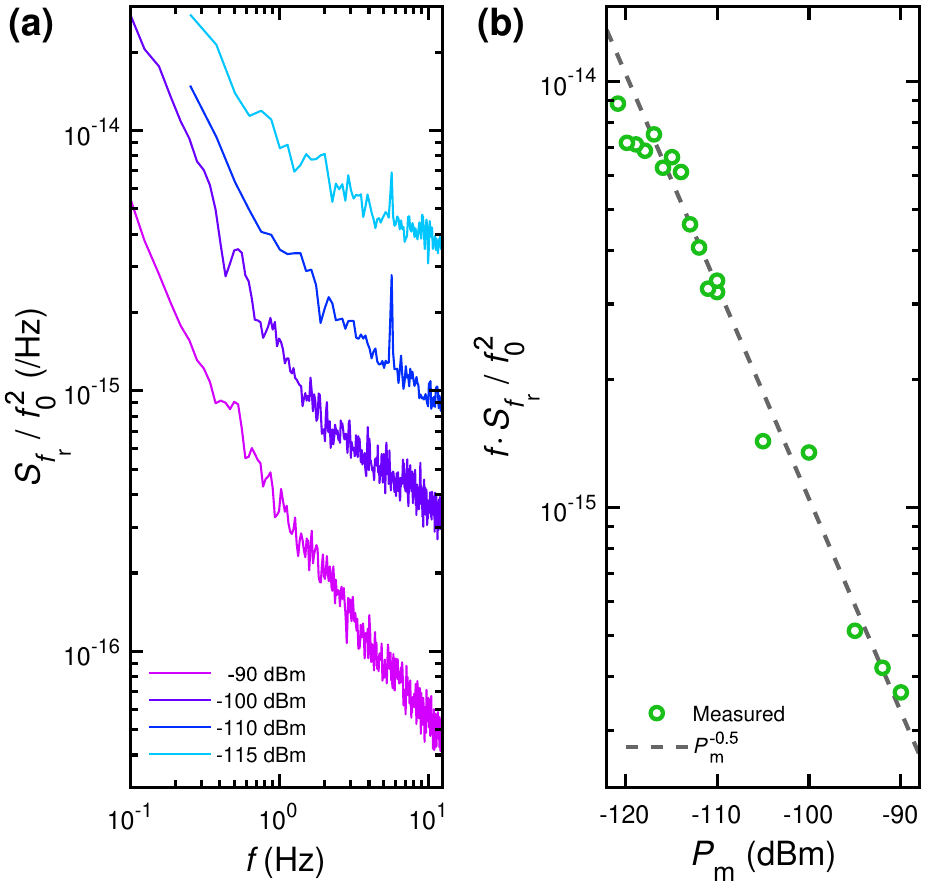}
    \caption{
    {\bf Frequency noise measured at different carrier powers} at the cryostat base temperature 10~mK.
    {\bf(a)} Raw scaled noise spectra $S_{f_\mathrm{r}}/f_0^2$ for measurement powers \num{-115}, \num{-110}, \num{-100} and \qty{-90}{dBm} ($\bar{n}=10^5$).
    {\bf(b)} Frequency fluctuations as a function of carrier power. The data were taken from the $1/f$ part by fitting raw spectra by $1/f$ and white noise. The amount of $1/f$ noise decreases with increased measurement power up to the largest carrier power.
    The dashed line indicates $P_\mathrm{m}^{\alpha}$ dependence with $\alpha = -0.5$.
    }
    \label{fig:Power_dep}
\end{figure}

The data in Fig.~\ref{fig:NoisePower_T}(a) were fit using Eq.~(\ref{eq:1_f}) with $A$ and $\gamma$ as free parameters (power is kept constant so that $\alpha$ can be omitted here). The obtained noise exponent $\gamma$ is displayed in Fig.~\ref{fig:NoisePower_T}(b) as a function of temperature. The data indicate clear temperature dependence in $\gamma(T)$ between 20 and 200~mK. The results on $\gamma(T)$ are similar to those obtained for flux noise in SQUIDs where magnetic impurities are supposed to dominate the $1/f$ noise \cite{Anton2013}.
 
Using the fit results for $A$ and $\gamma$, we find a pivot point for the noise traces at \qty[separate-uncertainty,separate-uncertainty-units = single]{0.8(0.5)}{Hz} (see Fig.~\ref{fig:NoisePower_T}(a)). Also, this pivoting feature is similar to findings in SQUIDs in Ref.~\onlinecite{Anton2013}. Fig.~\ref{fig:NoisePower_T}(c) depicts the noise power $f\cdot S_{f_\mathrm{r}}/f_0^2$ measured at 1, 10 and \qty{100}{Hz} from fitting to Eq.~(\ref{eq:1_f}). Near the pivot point, the noise power presents no clear dependence on temperature while clear increase is seen at 10 Hz and 100 Hz.

Frequency fluctuation spectra at different carrier powers are shown in Fig.~\ref{fig:Power_dep}(a). The noise spectra were averaged \numrange[range-phrase = --]{200}{900} times, with the number of averages decreasing with increasing power.  The largest measurement power $P_\mathrm{m}=$~\qty{-90}{dBm} yielded nearly $1/f$ type of noise while clear deviation due to white noise could be observed at smaller powers. In our measurements, the number of averages could be lowered with increasing $P_\mathrm{m}$ because the thermal background noise was found to decrease as $P_\mathrm{m}^{-1}$ while the $1/f$ noise decayed slower with $P_\mathrm{m}$ \footnote{The enhanced white noise here from the detection circuitry is due to the chosen large dynamic range of the system with only rather small gain before the diode detector.}.
Fig.~\ref{fig:Power_dep}(b) displays the observed decrease in $1/f$ noise with increasing carrier power. $P_\mathrm{m}^{-0.5}$ dependence is clearly observed from \num{-120} to \qty{-90}{dBm} carrier power.

	
	\section{Discussion}

Previous experiments have demonstrated that thin CoSi\textsubscript{2} films possess a substantial temperature-dependent kinetic inductance \cite{Mukhanova2023}. These experiments yielded for temperature dependence of the cavity frequency $f_0^{-1}\mathrm{d}f_0/\mathrm{d}T \simeq 0.2 T^3$~\unit{K^{-4}}. At $T=0.2$~K, this corresponds to $f_0^{-1} \mathrm{d}f_0/\mathrm{d}T \simeq 16 \times 10^{-4}$~\unit{K^{-1}}, which yields $\delta f_0 \sim 8 $~Hz for one \unit{\micro K} fluctuation \footnote{Our resistance bridge reading display approximately \qty{1}{\micro K} rms-noise at 1 Hz for the mixing chamber temperature. This noise, however, may be electronic in origin so it has to be considered as an upper limit estimate for temperature fluctuations.}. Such a fluctuation corresponds to approximately 10\% of the frequency fluctuations of 100 Hz observed at 1~Hz.

In SQUIDs, it has been argued that clusters of spins may dominate the flux noise. Ferromagnetic clustering could lead to large enough local fields that would yield time-varying dynamics of spins, interacting with the order parameter and with current-induced microwave flux, which could result in low frequency noise in the measured microwave transmission.  Typically, flux noise due to such clustering spins influence dissipation in the microwave cavity, and they lead to a characteristic increase of $Q$ as a function of drive power. Even though a complete microscopic model for the magnetic moments acting on the order parameter is missing, we may expect that a similar order parameter suppression may arise from the trapped flux present in our sample without magnetic shielding. Fluctuations in the order parameter lead to variation in kinetic inductance which is seen as frequency noise. 

One way to estimate the expected inductance noise magnitude is to consider the cavity as a long weak link for which the critical current $I_\mathrm{c}$ is related to the normal state resistance by the Ambegaokar-Baratoff formula: $I_\mathrm{c} R_\mathrm{n} \sim \Delta/e$. For a wire, this can be taken as valid for a section with length equalling the coherence length, amounting to $\xi(0)=$~\qty{90}{nm} at zero temperature \cite{Chiu2021}.
As the length of our $\lambda/4$ cavities is around \qty{4}{mm}, $\xi(0)=$~\qty{90}{nm} means that the total kinetic inductance of \qty{\sim 1}{nH} is built from $N_L \sim 4 \times 10^4$ sections with inductance of \qty{25}{fH}. The critial current $I_\mathrm{c}$ can be estimated using the equation for kinetic inductance $L_\mathrm{K}=\frac{2}{3\sqrt{3}}\frac{L_\mathrm{line}}{\xi}\frac{\hbar}{2e I_\mathrm{c}}$ \cite{Tinkham2002,Bezryadin2013} which yields $I_\mathrm{c} \simeq 5$~mA. As the sheet resistance $R_\mathrm{s}=$~\qty{2.4}{\ohm}, this supercurrent yields $I_\mathrm{c} R_\mathrm{n}^{\xi(0)} =$~\qty{0.1}{mV} where $R_\mathrm{n}^{\xi(0)}=$~\qty{0.02}{\ohm} is the resistance over one coherence length. This $I_\mathrm{c} R_\mathrm{n}^{\xi(0)}$ value is close to the energy gap in CoSi\textsubscript{2}.
 
To estimate $I_\mathrm{c}$ fluctuations, we need an approximate number of independent fluctuations in resistance. In CoSi\textsubscript{2}, $1/f$ resistance fluctuations at $f=$~\qty{1}{Hz} amount to an exceptionally small value: $\delta R/R \simeq 10^{-6}$ \cite{Chiu2017}. This is independent of length $L$ as long as $L < \ell_\mathrm{in}$, the inelastic coherence, which can be microns $\ell_\mathrm{in} \sim$~\qty{10}{\micro m} in normal metals at mK temperatures \cite{Giazotto2006}. Thus, the resistance fluctuations can be considered to be built from $N_\mathrm{R} \sim 400$ sections with $\delta R/R \simeq 10^{-6}$, which is reduced by $1/\sqrt{N_\mathrm{R}}$, yielding for the whole length $\delta R/R \simeq 5 \times 10^{-8}$; this gives a conservative estimate for the inductance noise. However, the number of independent inductive sections  $N_\mathrm{L}$ in the superconducting state contributing to $L_\mathrm{K}$ may be even larger. Assuming $\delta R_\mathrm{n}^{\xi(0)}/R_\mathrm{n}^{\xi(0)}\simeq  10^{-6}$ as the basic independent unit, we obtain an even larger reduction in noise, resulting in an expectation value of $\delta L/L \simeq 5 \times 10^{-9}$. This would yield $\delta f_0/f_0 \sim 3 \times 10^{-9}$, which is well comparable with typical frequency fluctuations in microwave cavities, mostly induced by two-level systems modifying the dielectric properties of the substrate and the surface of the conductor \cite{Burnett2014}. Our result at small power is \numrange[range-exponents = combine,range-phrase = --,range-units = single]{2e-8}{8e-8} at \qty{1}{Hz}, which is close to the upper estimate based on the normal-state resistance noise with 400 independent sections. Since the dissipation in our CoSi\textsubscript{2} cavities is dominated by other processes than TLSs \cite{Mukhanova2023}, it is unlikely that the two level states would dominate the noise. Further experiments are needed in order to distinguish 
whether the noise is due to the underlying scattering fluctuations inducing resistance noise in the normal state or due to some other process. 
 
	\section*{Acknowledgments}
 We are grateful to Visa Vesterinen and VTT Technical Research Centre of Finland for generous support with the Josephson traveling wave parametric amplifier and to Manohar Kumar, Shao-Pin Chiu, and Tero Heikkil\"a, for fruitful discussions, and Pei-Ling Wu and Shouray Kumar Sahu for help in the experiments. This work was supported by the Academy of Finland (AF) projects 341913 (EFT), and 312295 \& 352926 (CoE, Quantum Technology Finland), as well as Taiwan-Finland AF mobility grant 341884 (S.~S.~Yeh). The research leading to these results has received funding from the European Union’s Horizon 2020 Research and Innovation Programme, under Grant Agreement No.~824109 (EMP). Support from Jane and Aatos Erkko Foundation and Keele Foundation (SuperC project) is also gratefully acknowledged. The experimental work benefited from the Aalto University OtaNano/LTL infrastructure. J.~J.~Lin acknowledges the support by the National Science and Technology Council (NSTC) of Taiwan through Grant Nos.~110-2112-M-A49-015 and 111-2119-M-007-005. 
 S.~S.~Yeh is grateful to the support by NSTC of Taiwan through Grant No.~110-2112-M-A49-033-MY3 and the support by the Taiwan Ministry of Education through the Higher Education Sprout Project of the NYCU.

\section*{DATA AVAILABILITY STATEMENT}
 The data that support the findings of this study are available from the corresponding author upon reasonable request.

	\onecolumngrid


	\twocolumngrid
%
	
	\appendix

\end{document}